\documentclass[
    ,final            
  ]
  {aipproc}

\layoutstyle{6x9}

\begin{document}

\title{The X--ray afterglow of GRB030329 at early and late times}

\author{A. Tiengo}{
  address={Istituto di Astrofisica Spaziale e Fisica Cosmica --
              Sezione di Milano ``G.Occhialini'' (Italy)}
,altaddress={Universit\`{a} degli Studi di Milano,
            Dipartimento di Fisica (Italy)}
}

\author{S. Mereghetti}{
  address={Istituto di Astrofisica Spaziale e Fisica Cosmica --
              Sezione di Milano ``G.Occhialini'' (Italy)}
}

\author{G. Ghisellini}{
  address={INAF-Osservatorio Astronomico di Brera, Merate (Italy)}
}

\author{E. Rossi}{
  address={Institute of Astronomy,  Cambridge (UK)}
}

\author{G. Ghirlanda}{
  address={Istituto di Astrofisica Spaziale e Fisica Cosmica --
              Sezione di Milano ``G.Occhialini'' (Italy)}
}

\author{N. Schartel}{
  address={XMM-Newton Science Operation Center, ESA, Vilspa (Spain)}
}

\begin{abstract}
Thanks  to its extraordinary brightness, the X-ray afterglow of GRB030329
could be studied by {\it XMM-Newton} up to two months after the prompt
Gamma-ray emission. 

We present the results of two {\it XMM-Newton} observations performed 
on May 5 and 29, as well as an analysis of the {\it Rossi-XTE} data of 
the early part of the afterglow, discussing in particular
the stability of the X--ray spectrum
and presenting upper limits on the presence of X--ray emission lines.
\end{abstract}

\maketitle


\section{INTRODUCTION}

GRB030329 is an exceptional Gamma--ray burst for various reasons:
it  had a very large fluence of $\sim$10$^{-4}$ erg cm$^{-2}$ (30--400 keV, 
\cite{Ricker:2003}), in the top 1\% of all observed GRBs;
its redshift is  z=0.1685 (\cite{Greiner:2003},\cite{Caldwell:2003}), 
which makes it the second nearest GRB;
its optical transient was observed at magnitude 13 one hour after the
  explosion (\cite{Price:2003},\cite{Torii:2003});
it is the first GRB unambiguously associated with a supernova (\cite{Stanek:2003}, 
\cite{Hjorth:2003}).
The detailed studies in all wavebands made possible by the brightness of
  this event are yielding an unprecedented understanding on the jet
  structure, GRB energetics and circumburst environment.
In particular, the X-ray afterglow could be studied at late times, with a 
sensitivity which was not achieved for previous bursts.

The spectral shape and the time evolution of
the X--ray afterglow has been already reported and discussed in comparison with
preliminary measurements of the optical afterglow by \cite{Tiengo:2003}.

\section{STABILITY OF THE NON--THERMAL X--RAY SPECTRUM}

The first part of the afterglow of GRB030329 was studied with {\it Rossi-XTE}, which
observed it twice in the first 30 hours since the burst explosion.
For visibility constraints GRB030329 could not be observed by  {\it XMM-Newton}
until May.
The first XMM observation was carried out 37 days after the GRB and a second
one was done 23 days later.

All the available X--ray spectra of the GRB030329 afterglow are well 
fitted by an absorbed power law with photon index $\sim$2.2 and absorption fixed to the
Galactic value in that direction (N$_{H}$=2$\times$10$^{20}$ cm$^{-2}$). 
Such a non--thermal model is quite typical for X--ray afterglows, which are usually
observed within few days since the GRB explosion. However, it is remarkable 
that it can fit also the afterglow two months after the GRB, when its flux
has already decayed by $\sim$4 decades (see Fig.1).

\begin{figure}
  \includegraphics[height=.7\textheight, angle=-90]{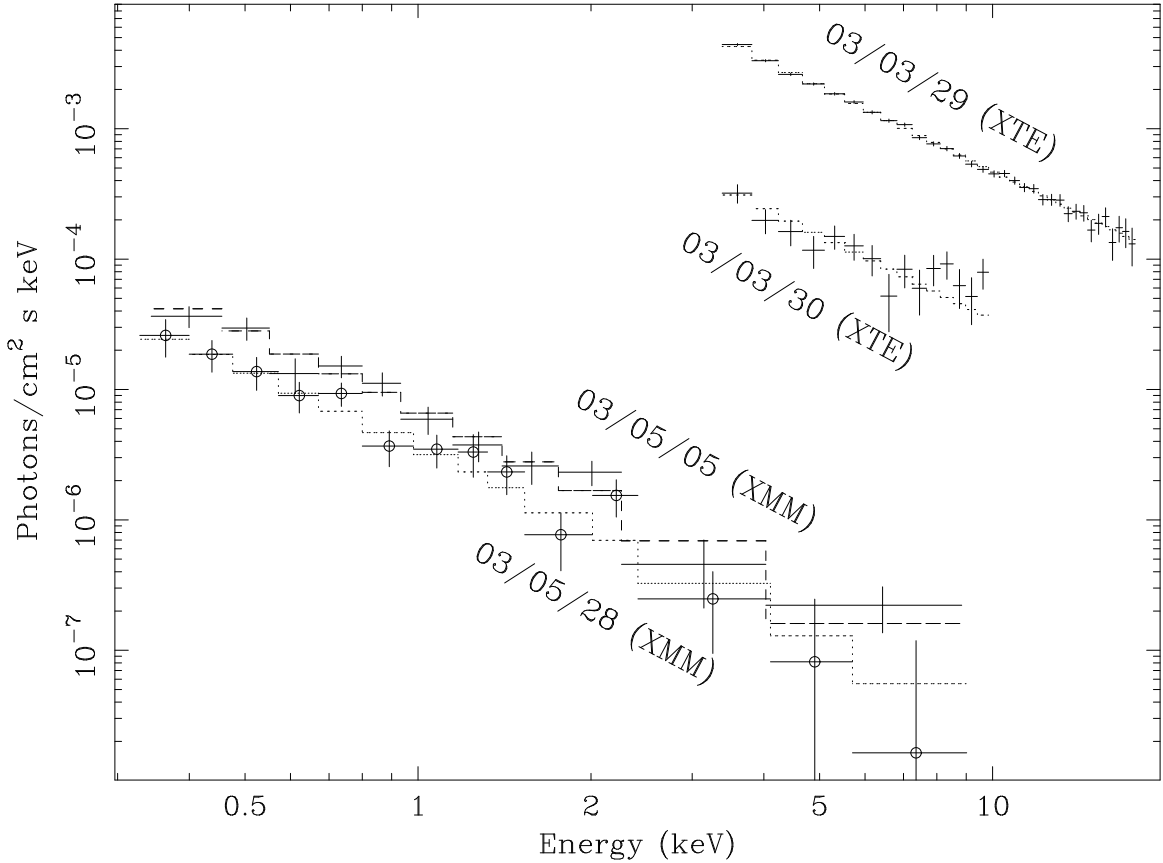}
  \caption{The X--ray spectra of the GRB030329 afterglow taken 5 hours, 30 hours 
({\it Rossi-XTE} data), 37 days, and 61 days ({\it XMM-Newton} data) since the burst 
explosion. These unfolded spectra are obtained fitting the count spectra with an absorbed  
power law model with N$_{H}$=2$\times$10$^{20}$ cm$^{-2}$ and $\Gamma$=2.2.}
\end{figure}

The high quality of the EPIC spectral data allows us to investigate the 
possible alternative of non--thermal spectra for the late afterglow. 
We find that a fit to the summed spectra
of the two {\it XMM-Newton} observations with a thermal
plasma model (MEKAL) with Solar abundances is not acceptable ($\chi^{2}/dof$= 62.6/30, 
see Fig.2). If the abundance is left free to vary, an 
acceptable fit is obtained (28.5/29) with kT=2.4$\pm0.6$ and a 3$\sigma$ upper 
limit on the abundance of Z$<$0.2.

\begin{figure}
  \includegraphics[height=.7\textheight, angle=-90]{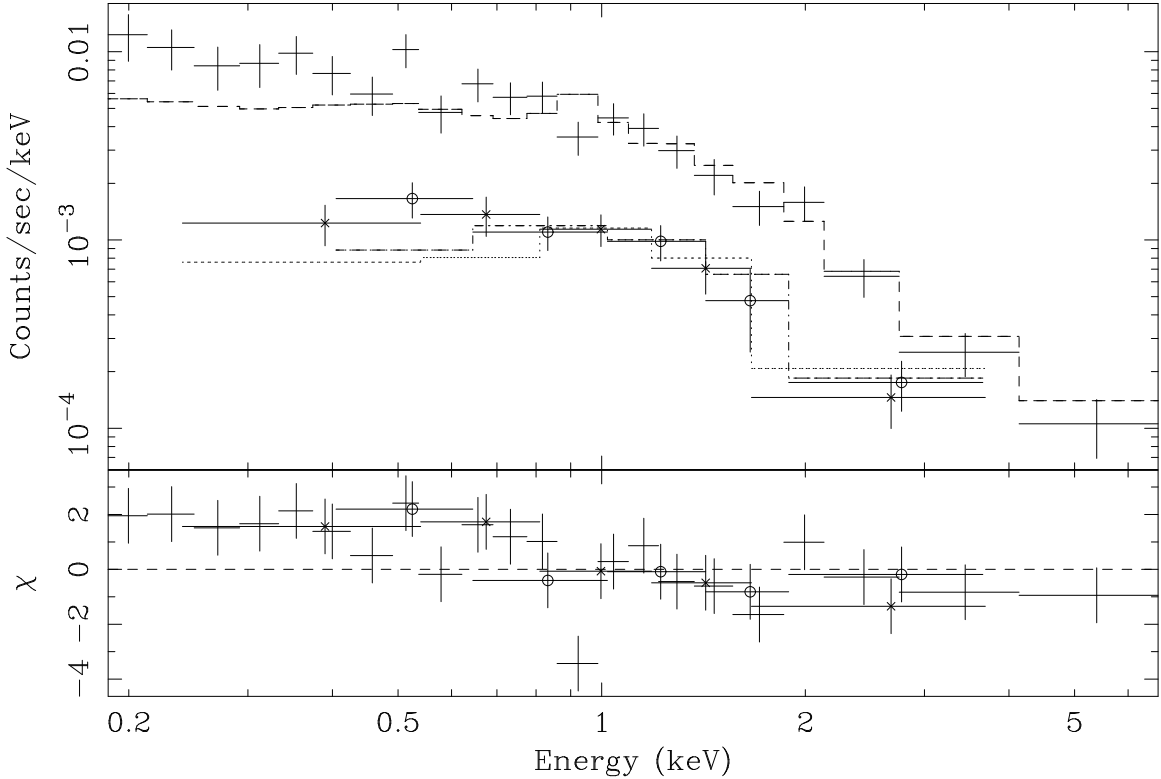}
  \caption{EPIC spectrum of the sum of the two
{\it XMM-Newton} observations fitted with a MEKAL 
model (kT=3.8 keV) with Solar abundances and redshift fixed at z=0.1685 
(absorption is fixed at the Galactic value N$_{H}$=2$\times$10$^{20}$ cm$^{-2}$)}
\end{figure}

\section{CONSTRAINTS ON EMISSION LINES}

No discrete spectral features have been significantly detected in any of the X--ray 
spectra of the GRB030329 afterglow. 

To derive upper limits on the presence of narrow emission lines in the late X--ray afterglow
of GRB030329, we fitted the EPIC MOS and PN spectra of the sum of the two {\it XMM-Newton} 
observations with a model consisting of an absorbed (N$_{H}$ fixed to the Galactic value) 
power law and a Gaussian emission line with $\sigma$=0. The line centroid was
fixed to a grid of values covering the 0.5--6 keV energy band and all the 3$\sigma$ upper limits on the
corresponding normalizations were computed. The maximum values of the corresponding
equivalent widths in different energy ranges are reported in Table 1. We consider them a reliable 
estimate of our sensitivity in detecting narrow emission lines in the EPIC data.

The detection of emission lines in the soft X--ray spectrum  has been reported 
in the afterglows of GRB011211 (\cite{Reeves:2002}), GRB030227 
(\cite{Watson:2003}), GRB020813, and GRB021004 (\cite{Butler:2003}). In most of
these cases the lines could be detected only during short time intervals in the early
phases of the afterglow. All the lines
were identified with transitions of rest--frame energies between 1.4 and 4.6 keV. 
For the redshift of GRB030329 their range corresponds to the 1.2--4 keV observed band.
The two lines detected with the highest
significance in all these cases (Si~{\small XIV} and S~{\small XVI}) have 
energies of 2.22 and 2.77 keV (1.9 and 2.4 keV at z=0.1685)
and had equivalent widths smaller than 600 eV. 
These results can be compared with our upper limits for the afterglow of GRB030329 (Table 1).

Due to the combination of low redshift, small interstellar 
absorption and high quality spectra, we can put stringent limits to the presence of 
emission lines with rest--frame energy much lower than in any other GRB afterglow. 

On the contrary, no significant information is obtained on the presence of a Fe-K line, 
which, at z=0.1685, is expected in the 5--6 keV band, where only few photons were 
collected in the EPIC instrument.




\begin{table}
\begin{tabular}{crcr}
\hline
& \tablehead{1}{r}{b}{Equivalent width}\\
\hline
0.5--1 keV & $<$120 eV \\
1--1.5 keV & $<$150 eV\\
1.5--2 keV & $<$400 eV\\ 
2--3 keV & $<$700 eV\\
3--5 keV & $<$2000 eV\\
5--6  keV & $<$2800 eV\\
\hline
\end{tabular}
\caption{3$\sigma$ upper limits on emission lines}
\label{tab:a}
\end{table}




\bibliographystyle{aipproc}   

\bibliography{mereghetti_sandro_1}

\IfFileExists{\jobname.bbl}{}
 {\typeout{}
  \typeout{******************************************}
  \typeout{** Please run "bibtex \jobname" to optain}
  \typeout{** the bibliography and then re-run LaTeX}
  \typeout{** twice to fix the references!}
  \typeout{******************************************}
  \typeout{}
 }

\end{document}